\documentclass[12pt]{article}

\usepackage{graphicx}
\usepackage{amsmath,amssymb,color,times,float,amsthm}
\usepackage{subfig,dcolumn,bm,caption,apacite}
\usepackage{mathptmx}      

\definecolor{background-color}{gray}{0.98}

\begin{document}

\pagestyle{empty}
\begin{center}
{\bf \large Adversarial Machine Learning for Cybersecurity \\[0.05in] and Computer
  Vision: Current Developments and Challenges} \\[0.15in]

Bowei Xi \\
Department of Statistics, Purdue University\\
Email:~xbw@purdue.edu 
\end{center}

\paragraph{Abstract:}
We provide a comprehensive overview of adversarial machine
learning focusing on two application domains, i.e.,
cybersecurity and computer vision. Research in adversarial
machine learning addresses a significant threat to the wide
application of machine learning techniques -- they are vulnerable
to carefully crafted attacks from malicious adversaries. For
example, deep neural networks fail to correctly classify 
adversarial images, which are generated by adding imperceptible
perturbations to clean images. 
We first discuss three main
categories of attacks against machine learning techniques --
poisoning attacks, evasion attacks, 
and privacy attacks. Then the corresponding defense approaches
are introduced along with the weakness and limitations of the
existing defense approaches. We notice adversarial samples in cybersecurity and
computer vision are fundamentally different. While adversarial
samples in cybersecurity often have different
properties/distributions compared with training data, adversarial
images in computer vision are created with minor input
perturbations. This further complicates the development of robust
learning techniques, because a robust learning technique must
withstand different types of attacks.


\section{Introduction} 
\label{sec:intro}

\begin{figure}[b!]
\centering
\begin{verbatim}
I've been waiting for your reply
Tristan <xxx@verizon.net>
To:xxx@yahoo.com
Feb 6 at 6:41 PM
Hey Alice,

I have been trying to get in touch with you
I am inviting you on board to the system I got going a few weeks ago.
The first participants have seen a reasonable amount of success.
Your invite is attached, the password is 815452
\end{verbatim}\vspace{-0.8cm}
{\color{red}\begin{verbatim}! Attachments cannot be downloaded.\end{verbatim}} \vspace{-1cm}\begin{verbatim}Learn More\end{verbatim}\vspace{-0.5cm}
\caption{\label{fig:spam-email} A spam email sent on Feb. 06
  2019, and received in Yahoo Spam folder}
\end{figure}
There are two main branches of adversarial machine learning
research. One branch actively designs new attacks to defeat
existing machine learning algorithms and machine learning based
systems. At the same time, the other branch 
aims to significantly improve the
capabilities of machine learning techniques facing
attacks. 

Attacks against machine learning based systems were
observed first in cybersecurity, e.g., \cite{advl-security-2013}. One example of
adversarial attacks is 
network intrusion. Attackers actively scan the systems to discover new 
vulnerabilities in network devices, and to compromise vulnerable
hosts/systems. Machine learning techniques were applied to
identify the discriminant features of malicious
network traffic, e.g., \cite{intrusion-featureselect-2003}, and  
classify legitimate network traffic from malicious traffic, e.g.,
\cite{intrusion-mining-1998,intrusion-NN-1998,intrusion-NN-SVM-2002,intrusion-ML-review-2009}.
As noted early on, e.g., \cite{computer-attacks-2002}, attack tools are
updated frequently to avoid detection. Hence it is not a trivial
task to identify the malicious traffic since the signatures of
the malicious traffic change suddenly. Another example is 
spam. Spam filters on email
servers serve as classifiers that identify spam emails. Spam filters either
directly block the spam emails or place them in a separate spam 
folder. Although people have been combating spam emails for over 20
years, e.g., \cite{spam-1998}, spam emails are still a nuisance
today.  Spammers frequently
change how spam emails are composed to avoid detection. Spammers
in early 2000s mis-spelled 
the words that could easily be identified, such as writing
``Pharmaceutical'' as ``Phaxrrmaceutical'', and inserted good
words, i.e., words often observed in legitimate emails, into spam
emails \cite{lowd2005good}. 
 Figure~\ref{fig:spam-email} shows a spam email
received in Yahoo Spam folder in Feb, 2019. The entire content of
this recent spam email is similar to legitimate emails.

\begin{figure}[tb]
\centering{
\includegraphics[width=3in]{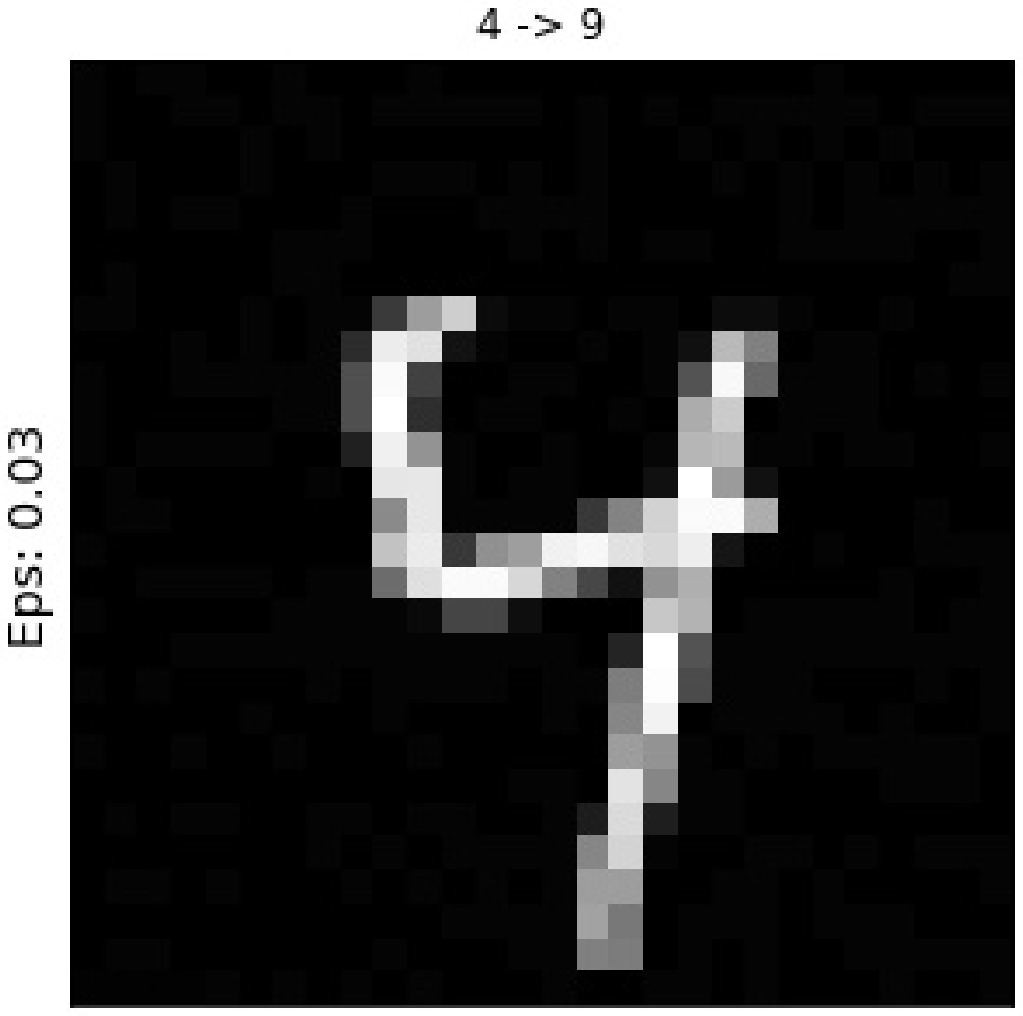}
\hspace{0.1in}
\includegraphics[width=3in]{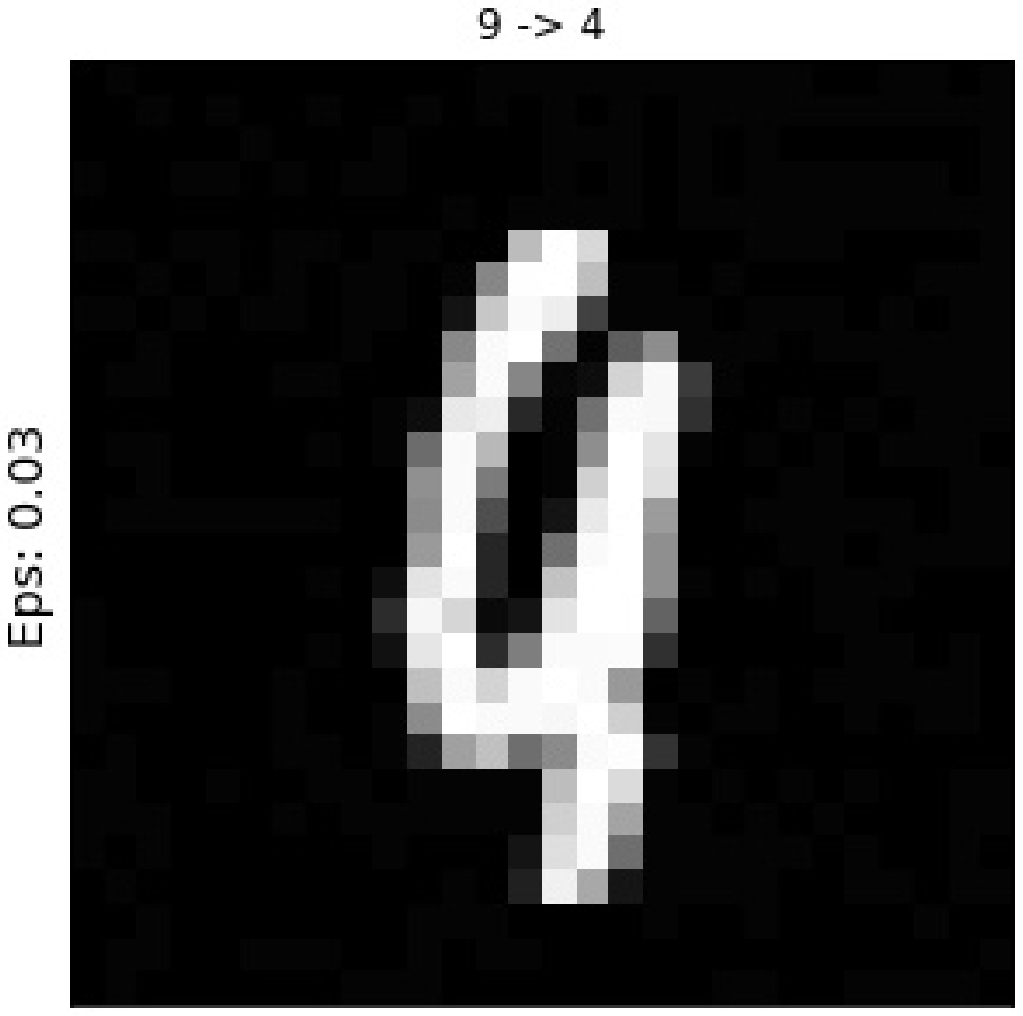}
}
\caption{Two adversarial samples. Left: A handwritten 4
  is misclassified as 9; Right: A handwritten 9 
  is misclassified as 4.} 
\label{fig:advl-image}
\end{figure}

Recently adversarial samples are designed in computer
vision to break deep neural networks (DNN). Although DNNs are capable of
performing complex tasks, they are shown to be vulnerable to
attacks. Adversarial images created with minor perturbations are
misclassified by DNN, as illustrated in
Figure~\ref{fig:advl-image}. 

Adversarial attacks against DNN draw significant attention due
  to the wide use of DNN in critical tasks, such as autonomous driving systems and partly
  automated vehicles. For example, Tesla has a on-board computer that ``runs
  the new Tesla-developed neural net for vision, sonar and radar
  processing software ... ... Built on a deep neural network, Tesla
  Vision deconstructs the car's environment at greater levels of
  reliability than those achievable with classical vision
  processing techniques.'' 
\footnote{https://www.tesla.com/autopilot (last verified on
  03/01/2020)} 
On the other hand, in one Tesla fatal crash, ``The car veered off the road due to
limitations of the Tesla Autopilot vision system’s processing
software to accurately maintain the appropriate lane of travel.'' 
\footnote{https://www.washingtonpost.com/local/trafficandcommuting/deadly-tesla-crash-tied-to-technology-and-human-failures-ntsb-says/2020/02/25/86b710bc-574d-11ea-9b35-def5a027d470\_story.html
  (last verified on 03/01/2020)} 
Such high profile failure of the hugely popular deep learning
technique shows the importance of adversarial machine learning
area. Researchers must fully understand the vulnerabilities
of the learning techniques and consequently robustify the existing learning
techniques for them to be used reliably in real life tasks.

There are different types of adversarial samples. Adversarial
images are samples with minor perturbations added to the clean
images, whereas adversarial samples in cybersecurity have very
different properties/distributions compared with the training samples. 
Furthermore, adversaries seek
to learn information about the training data points with
sensitive information through a learning model, causing a
privacy leakage. 

Hence machine learning techniques need to significantly improve their
generalization capability to 
properly handle the adversarial samples created with only minor
perturbations, as observed in computer vision. Learning techniques
also need new capabilities to 
identify adversarial samples that change quickly and have very
different properties, as observed in
cybersecurity. Meanwhile, learning 
models with complex structure, such as DNNs, may still suffer privacy leakage. 
It is an urgent and important task to develop secure and robust
machine learning techniques that can withstand different types of
malicious attacks. Currently there are different
approaches. We observe ideas are converging from different
communities, such as cybersecurity, 
computer vision, theoretical machine learning, and artificial
intelligence, to achieve this goal.   

In this review article, we
introduce different types of malicious attacks, the existing 
strategies to robustify and secure machine learning techniques, and the 
challenges faced in adversarial machine learning area. 
The rest of the paper is
organized as follows. Section~\ref{sec:attack} discusses the
adversarial attacks; Section~\ref{sec:defense} introduces the
current defense approaches; Section~\ref{sec:conclude} discusses 
the challenges that remain to be addressed and concludes this article.  

\section{Adversarial Attacks}
\label{sec:attack}

\begin{figure}[tb]
\centering{
\includegraphics[width=5in]{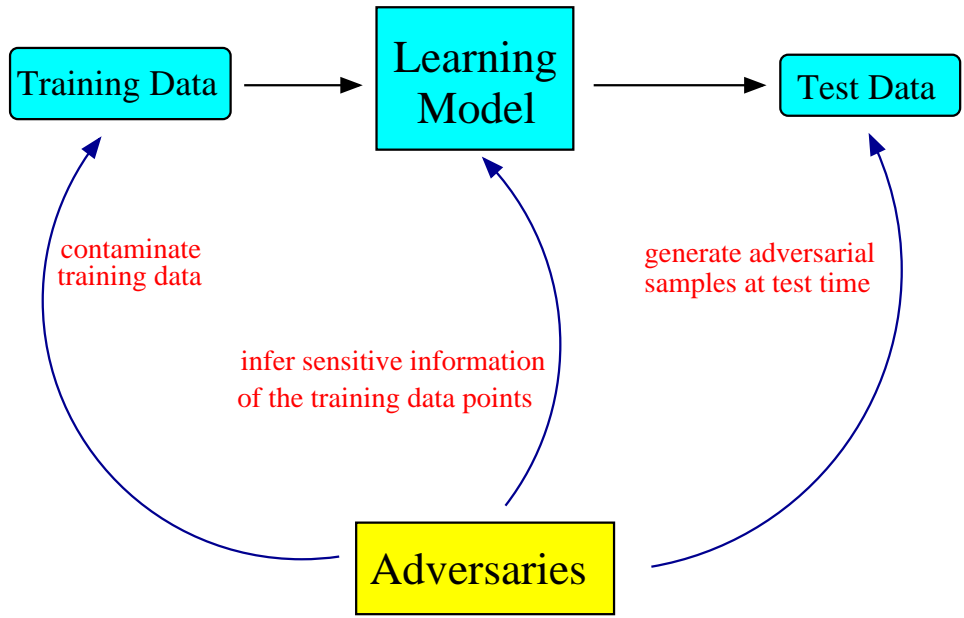}
}
\caption{Illustration of poisoning attack, evasion attack, and
  privacy attack } 
\label{fig:attack-3types}
\end{figure}

Machine learning techniques are widely used in different
applications. We observe many exciting developments of new
techniques, such as the success of DNN in computer vision tasks.
Unfortunately there are a growing number of vicious attacks against 
learning models too. 
In this section we introduce three common forms
of attacks. As illustrated in Figure~\ref{fig:attack-3types},
adversaries can contaminate the training data to make a classifier ineffective 
(poisoning attack),
create adversarial samples at test time to evade detection
(evasion attack), and
infer sensitive information about the training data through a
learning model (privacy attack). 
\subsection{ Poisoning Attack}

Poisoning attacks aim to contaminate the training dataset and
cause a learning model to make costly mistakes at the test
time. Poisoning attacks, as well as evasion attacks, can be
either targeted or non-targeted. A targeted attack causes a
learning model to make a specific mistake, e.g., misclassifying a
certain type of legitimate emails. A non-targeted attack reduces
the overall accuracy of a learning model. 
Earliest examples of poisoning attacks are those against a spam
filter \cite{poison-spam-2008,attacks-2010,poison-2011}, against
an anomaly detection algorithm 
\cite{poison-http-2009,poison-anomaly-2010}, and against
intrusion detection systems \cite{intrusion-poison-survey-2013}.

\begin{itemize}
\item \cite{poison-spam-2008,attacks-2010,poison-2011} considered a
scenario where a spam filter were periodically retrained, and
spammers sent emails containing words frequently observed in
legitimate emails. Although the emails from spammers were labeled
correctly as spam, by including them in the retraining
process, they caused a spam filter to misclassify legitimate emails. 

\item \cite{poison-http-2009,poison-anomaly-2010} provided a
quantitative strategy to design and analyze the effectiveness of
poisoning attacks against centroid based anomaly detection
algorithm, which used the distance from the mean of the training
data to define anomalies. The effectiveness of a poisoning attack
was measured by the relative distance of how much the center was
moved by injecting new attack data points into the training
dataset. Upper bounds were established under partial control
and full control of the training data.   

\item Adversarial attacks in cybersecurity are generated with a good
understanding of the computer systems and their 
security policies. Poisoning attack is one of the six attack
categories against intrusion detection systems, as discussed  
in \cite{intrusion-poison-survey-2013}.   

\end{itemize}

Meanwhile we also witness poisoning attacks against specific
learning models, including support vector machine (SVM), LASSO and DNN. 

\begin{itemize}
\item \cite{poison-svm-2012} introduced an
algorithm to generate a poisoning attack sample $x^*$ against
SVM in a two class scenario. Starting from a randomly
selected sample with flipped label, the algorithm computed the
gradient of the hinge loss function  $\frac{\partial L(x^*)}{\partial w}$
and iteratively searched for a poisoning attack sample to
maximize SVM's validation error.  

\item \cite{poison-lasso-2015} introduced poisoning attack against
penalized regression models, including LASSO and elastic
net. The algorithm generated a poisoning attack sample $x^*$  
through maximizing the objective function, 
$max_{x^*} \left(\frac{1}{n}||\hat{Y}-\hat{W}'X-\hat{b}||_2^2
  +\lambda\times p(W)\right)$, 
where $p(W)$ is the corresponding penalty term for LASSO or
elastic net. The generated attack samples rendered the
associated feature selection process unstable. 
\cite{poison-lasso-2015} showed LASSO selected nearly random
variables when less than 5\% of generated
poisoning attack samples were added to the training data.  

\item \cite{dnn-troj-2017} discussed the danger of having a
``backdoor'' in a DNN, a.k.a. Trojan attack. A DNN trained by a
  malicious adversary 
behaved normally for most of 
the samples, and produced an error only when a backdoor trigger was
present. For example, a sticker posted on a stop
sign caused it to be misclassified.   
\end{itemize}

\subsection{  Evasion Attack}
Evasion attacks are the most popular form of attacks. Often
malicious adversaries do not have access to the training
process. They generate adversarial samples to cause
misclassification by a classifier or to evade detection by a
learning algorithm at the test time. 
Spam emails, malwares, and network intrusion traffic observed at the test
time, e.g., \cite{malware-srndic-2013,malware-papernot-2017}, are
examples of evasion attacks in cybersecurity. 

More recently, in computer vision, many evasion  
attack algorithms are developed to generate adversarial samples
that lead to misclassification by DNNs, starting with the first paper
published in 2014 \cite{attack-dnn-1st-2014}. Attacks against DNN
can be either digital or physical. A digital attack directly
add minor perturbation to the input to a DNN, i.e., an image, whereas a physical
attack creates minor perturbation on a physical object which
leads to DNN's failure to properly recognize the physical
object. There are few physical attacks compared with a large
number of digital attacks. We focus on digital attacks in this
section and discuss physical attacks in
Section~\ref{sec:future}. 

Several survey articles are published providing the
detailed timeline of adversarial 
attacks against DNNs, e.g.,
\cite{survey-attack-defense-DNN-2019,survey-attack-2018,advl-10year-2018}.
In 2017, a NIPS competition organized by Google Brain attracted
many researchers to develop new adversarial attack algorithms and
effective defenses against attacks. 
The winning teams' results are documented in \cite{nips2017competition}. 
Due to the limited space, here we provide a summary list of
digital evasion attacks against DNNs \footnote{Code for many 
attack algorithms and some defenses can be found on GitHub, e.g.,
\cite{IBM-toolbox,cleverhan}.}.    

Attacks can be categorized based on attacker's knowledge of the
learning model. White box attack refers to attacker having a complete knowledge
about the DNN under attack, including the model structure and
all the parameter values. Black box attack means attacker does not
have access to a DNN's internal structure. Often in black box
attack, an attacker sends queries to probe the target
DNN. The attacker then uses the queries labeled by the target DNN to
train a substitute model and generates adversarial samples
against the substitute model. Majority of the digital attacks against
DNN are white box attacks. Below we introduce the most notable white
box attacks and black box attacks. 

\begin{itemize}

\item First we introduce the white box attacks.

\begin{itemize}

\item {\bf L-BFGS:} Vulnerability of DNN was first reported in
  \cite{attack-dnn-1st-2014}: Adding certain non-random
  imperceptible perturbation $\delta$
  to an image $x$ can cause a DNN $C$ to misclassify the adversarial 
  image $x^a=x+\delta$. The paper also observed the transferbility of
  adversarial images -- the specific perturbation $\delta$ can cause a
  different DNN $C^*$ to misclassify the same adversarial image
  $x^a$. For a given image $x$ and a target class label $t$,
  $\delta$ was obtained by solving a box constraint optimization
  problem as follows.   
$$
\textrm{min}~||\delta||_2,~~s.t.~~ C(x+\delta)=t,~~x+\delta\in [0,1]^m
$$  
The L-BFGS method was used to find an approximate solution.

\item {\bf Fast Gradient Sign Method (FGSM)} and its variations: Let $J(x,y)$ be the
  cross-entropy cost function and $y$ be the true label of a
  clean image $x$. Using the sign of the gradient of the cost
  function, \cite{advl-train-2015} generated adversarial image $x^a$ as 
$$ x^a = x + \epsilon ~\textrm{sign}(\nabla_x J(x,y)).$$
Compared with L-BFGS, FGSM lowered computation cost to
generate an adversarial image. FGSM is a one-step attack. There
are variations of FGSM, including improved one-step attacks or
iterative attacks. For example, \cite{attack-fgsm3-2016} produced more diverse
adversarial images by using the scaled gradient instead of the
sign of the gradient. \cite{advl-train-ensemble-2018} proposed a small random
perturbation in the one step attack.  
\cite{attack-fgsm2-Kurakin-2017} suggested
adversarial images can be generated using the predicted labels in
FGSM instead of the true labels. 
Instead of a simpler one step attack, adversarial
samples can be generated by iteratively following the direction
of the gradient while clipping the generated image to stay inside the
$\epsilon$ ball of the clean image $x$, i.e., the Basic Iterative Method (BIM) in
\cite{attack-BIM-physical-2017}. 
Projected Gradient Descent (PGD) attack \cite{attack-PGD-2018}
showed that BIM with random starting points in the $\epsilon$
ball yielded stronger adversarial samples. 
\cite{attack-moment-ensemble-2018}
further used momentum in iterative FGSM to increase attack strength and
maintain transferbility of adversarial images. The update with
momentum is the following. 
$$
h_{t+1} = \alpha h_t + \frac{\nabla_x J(x_t,y)}{||\nabla_x
  J(x_t,y)||_1},~~~
x_{t+1} = x_t + \epsilon ~\textrm{sign}(h_{t+1}). 
$$
To attack an ensemble of DNNs, \cite{attack-moment-ensemble-2018} computed the
cross-entropy $J(x,y)$ by averaging the logits of DNNs (i.e., output from the layer
before softmax) in the ensemble, then followed the direction of the 
gradient with a momentum to attack the
ensemble. 
 
\item {\bf Carlini and Wagner (C\&W)}: C\&W attack
  \cite{attack-carliniL2-2017} modified the
  objective function and used a different optimizer compared with
  L-BFGS attack. Their $L_2$ attack is 
  sufficiently strong to bypass a number of detection and defense
  methods. They solved the following box constraint optimization problem to find
  an adversarial perturbation $\delta$.
$$
\textrm{min}_{\delta} \left(~||\delta||_2 + c~\textrm{max}(~\textrm{max}_{l\neq o}(Z(x+\delta)_l)-Z(x+\delta)_o,-k~)~\right),~s.t.~~~x+\delta\in [0,1]^m,
$$
where $Z(x)_o$ is the output of the softmax for class $o$.  After
a change of variable, $\delta=\frac{1}{2}(tanh(w)+1)-x$, Adam
optimizer was used to search for $w$.  
Following the line of C\&W attack, Elastic-net attacks to DNNs (EAD) in
  \cite{attack-elasticnet-2018} added an elastic 
  net type of penalty term in the optimization problem to search
  for adversarial perturbations. EAD's $L_1$ attack achieved comparable
  performance as the C\&W $L_2$ attack.  

\item {\bf Jacobian-based Saliency Map Attack (JSMA)}: JSMA
computed the Jacobian of
  either the logits as in \cite{attack-saliency-map-2016} or the
  outputs of the softmax as in \cite{distill-nn-2016} for a
  clean image $x$, and built a saliency map. Based on the
  saliency map, JSMA iteratively
  chose a pixel to change in each step to increase the
  likelihood of the target class for an adversarial image.  

\item {\bf DeepFool}: \cite{attack-deepfool-2016} introduced an
  iterative attack algorithm to efficiently search for
  adversarial samples. At each iteration, a classifier $C(x_t)$
  was linearized at the current point $x_t$ and an update was
  computed as the scaled gradient with respect to the linearized
  classifier. \cite{attack-universal-2017} leveraged the DeepFool
  attack to find universal adversarial perturbations for almost
  all the clean images to fool a classifier. 

\item {\bf Adversarial Transformation Networks (ATN)}: 
  \cite{attack-ATN-2018} trained a neural network that modified a
  clean image into an adversarial sample to fool a target network
  or an ensemble of networks. ATN was trained by minimizing a
  loss function which balanced the loss on the input image $x$
  and the output of the target network. ATN can be used for both white
  box and black box attack. 
\end{itemize}

\item Next we introduce several black box attacks. 

\begin{itemize}

\item {\bf Zeroth Order Optimization (ZOO)}: ZOO
  \cite{attack-zoo-2017} changed the loss function in C\&W attack
  to 
$$\textrm{max}(~\textrm{max}_{l\neq o}(\log{F(x)_l}-\log{F(x)}_o,-k~),$$ 
where $F(x)$ is the
  output of a DNN. ZOO avoided directly computing the
  gradient. Instead ZOO approximated the gradient using the
  symmetric difference quotient method, with an increased
  computation cost. The 
  knowledge of a DNN's network structure was not required to
  approximate the gradient. 

\item {\bf OnePixel}: OnePixel attack \cite{attack-onepixel-2019}
  changed the value of only one pixel of a clean image to fool a
  DNN with differential evolution using only the predicted
  outputs from a DNN without knowing its network
  structure. OnePixel showed DNN is even vulnerable to very low
  dimension attack with limited information. 

\item \cite{attack-blackbox-papernot-2017} used the probing
  approach to train and generate adversarial images against a
  local substitute model. With 800 
  queries sent to Amazon Web Services and Google Cloud
  Prediction, most of the adversarial samples were misclassified
  by the target models hosted by Amazon and
  Google. \cite{attack-blackbox-ensemble-2017} generated a
  targeted attack against $k$ networks in the white box fashion,
  and showed the adversarial images can transfer to an additional
  black box network. \cite{attack-blackbox-2018} further
  developed a black box attack under limited queries and partial
  output knowing only the top $k$ class
  probabilities. Unlike previous attacks against image level
  classifier, \cite{attack-yolo-patch-2019} developed a black box
  attack against state-of-the-art object detectors. 
 
\end{itemize}

\end{itemize}

\subsection{  Privacy Attack}

Protecting data confidentiality has long been an important
research area. The research on privacy leakage and
  privacy preserving techniques preceded adversarial machine
  learning. Mostly it remains a separate
  research area. This review article includes privacy attack as a
third attack because the very recent privacy attacks targeting DNNs,
which points to a different type of DNN vulnerability. 
Recently there
are two notable privacy attacks -- model inversion attacks and
membership inference attacks -- 
designed to obtain sensitive training data information based on the 
outputs from complex learning models such as DNNs. Similar to
evasion attacks, privacy attacks can be either 
white box or black box, i.e., with or without access to the
internal structure and the parameters of the target learning model.  
A brief
review of different privacy preserving approaches as well as the
recent attempts to defend DNNs from privacy attacks are in Section~\ref{sec:privacy}.
 
\begin{itemize} 

\item {\bf Model Inversion Attacks}:
  \cite{model-inversion-attack-2015} designed two model inversion
  attacks against face recognition systems. The first attack can
  reconstruct a face image based on the person's unique label
  produced by a face recognition system. The second attack can
  obtain a clean image from a blurred image through attacking a
  system, thus obtaining the identity of the victim.  
 \cite{model-inversion-attack-2016} provided a formal description
 of model inversion attacks in black box scenario and white box
 scenario. It also suggested model invertibility was related to the
 influence/stable influence of Boolean functions. 

\item {\bf Membership Inference Attacks}: 
A basic membership inference attack discovers whether a data
point belongs to a learning model's training data, given only black box
access to the model \cite{member-infer-attack-2017}. 
They used several shadow models
trained on synthetic or noisy data to determine whether a data
point belonged to the training data or not. They successfully
designed membership inference attacks against Google
Prediction API and Amazon ML.
\cite{member-infer-attackEnsemble-2019} showed that when several
trained models periodically shared updates, the process can be
exploited to infer whether a data point was used in
training. They also suggested DNN's
complex structure made it vulnerable to membership inference
attacks under collaborative learning setting. 
\end{itemize}

\section{\sffamily \Large Robust and Secure Learning Strategies}
\label{sec:defense}

The concept of malicious noise and robustness of learning
algorithms including neural networks were studied as early as
1985. \cite{malicious-Valiant-1985} introduced a distribution
free model, i.e., the probably approximately correct (PAC) model,
which can tolerate a low rate of malicious noise.
\cite{robust-kearn-1993} established that the upper
bound of malicious error rate which can be tolerated by a learning
algorithm $A$ with accuracy $1-\epsilon$ is
$\frac{\epsilon}{1+\epsilon}$. \cite{robust-kearn-1993} further
constructed perceptron based PAC algorithms with malicious noise
tolerance.  Then \cite{defense-poison-boost-2001} built perceptron based PAC
algorithms with much higher level of malicious noise tolerance.     
\cite{zerosum-teo-2008} developed learning algorithms
that incorporated invariant transformations as a form of robust
learning procedure. 

Meanwhile researchers hope to
understand the baffling phenomenon of
adversarial perturbations that cause DNN to make mistakes. 
\cite{attack-dnn-1st-2014} suggested the highly non-linear nature
of DNN made it vulnerable, whereas \cite{advl-train-2015}
mentioned it was the locally linear structure which made DNN
vulnerable. It is generally agreed that DNNs with larger capacity
in terms of number of parameters are more robust, e.g.,
\cite{attack-fgsm2-Kurakin-2017,attack-PGD-2018}. 

In addition, recently researchers proposed various
  metrics to evaluate DNN's robustness and predictive uncertainty.
\cite{evaluate-classifier-2014} developed an algorithm to
empirically evaluate a classifier's performance under simulated
attacks.  \cite{robust-extreme-2018} proposed a robustness metric for DNN
based on extreme value theory, i.e., Cross Lipschitz Extreme Value for nEtwork
Robustness (CLEVER). 
\cite{attack-deepfool-2016} used the estimated minimum amount of
perturbations needed for an attack to fool a DNN as a measure of
DNN's robustness.  
\cite{property-advl-2019} investigated the relationship between
dimensionality and the robustness of a classifier. 

\cite{uncertainty-deep-ensemble-2016} used deep ensembles based
on random initializations and adversarial training
to obtain predictive uncertainty estimates. 
\cite{MonteCarlo-dropout-2016} used Monte Carlo dropout on test
samples to estimate predictive uncertainty. 
\cite{uncertainty-2018} used a Bayesian approach to evaluate DNN
uncertainty caused by distribution shift between training and
test samples. 

It is an ongoing effort to secure machine learning models. 
Here we introduce several approaches to improve the
performance of learning techniques in adversarial environment. 

\subsection{  Game Theoretic Approaches}

\begin{figure}[tb]
\centering{
\includegraphics[width=3.5in]{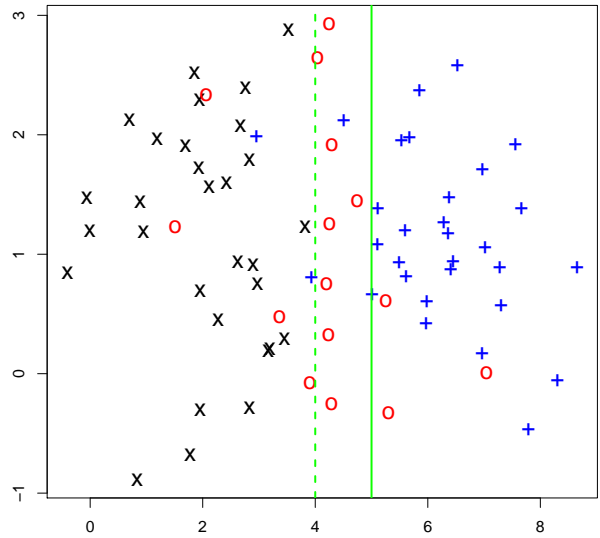}
}
\caption{A standard classification boundary (dashed line) vs. a
  conservative boundary (solid line)  } 
\label{fig:svm}
\end{figure}

Game theory provides a useful tool to model and understand the
interaction between the defender and the adversaries. 
Facing adversarial attacks, one strategy
  is to explore the robustness and accuracy trade-off when 
  building learning algorithms. An 
equilibrium solution can be utilized to construct robust learning
algorithms, which make conservative decisions for the current
data but have persistent good performance against potential
future attacks.  The learning algorithm's decision boundary depends on the
game that is used to model the interaction between defense and attacks.
The minimax solution in a zero-sum game leads to a conservative strategy
when the adversarial samples' property changes abruptly, whereas a
mixed equilibrium strategy suggests a randomized defense against
minor adversarial perturbations. Figure~\ref{fig:svm} is a conceptual
plot illustrating a conservative decision boundary. Sometimes a
minimax solution can be too conservative. Instead, a Stackelberg
game offers a less conservative equilibrium solution.   
To search for an equilibrium is often computationally
expensive. However, game theory offers important insights towards
resilient algorithms against malicious adversaries. Next we
discuss zero-sum game and Stackelberg game based robust learning
solutions.

\begin{itemize}
\item A two player zero-sum game where players take
  simultaneous actions is formulated as follows. Let $D$ be the
  defender and $A$ be the adversary. Let $U_D$ and $U_A$ be their
  corresponding utilities. Let $(d,a)$ be a pair of strategies,
  for the players $D$ and $A$ respectively. In a zero-sum game,
  $U_A(d,a)=-U_D(d,a)=U(d,a)$. For example, $d$ can be a
  classifier built by defender $D$ and $a$ can be a
  transformation of malicious objects by the adversary
  $A$. The Nash equilibrium is the optimal solution in a game
  where no player has the incentive to unilaterally change its
  strategy. The minimax solution $(d_e,a_e)$ is a Nash equilibrium in a two
  player non-cooperative game, defined as follows.
$$
U(d_e,a_e) = \textrm{min}_d ~ \textrm{max}_a~U(d,a) . 
$$ 
Hence at the equilibrium the defender $D$ wants to minimize its worst case loss. 

  \begin{itemize}
  \item \cite{zerosum-domingos-2004} applied a game theoretic
    approach to spam detection with Naive Bayes as the underlying
    classifier. The framework allowed a classifier to
    automatically adjust its classification rule facing evolving
    manipulations from an adversary. 
 
  \item \cite{zerosum-Lanckriet-2002} studied the binary
    classification problem. For a class, an ellipsoid shaped
    region was defined using the mean and the variance-covariance
    matrix. They assumed arbitrary sets of data points can be
    drawn from their respective ellipsoids. A robust linear
    classifier was proposed based on a minimax solution given
    this bounded data uncertainty. 
   
  \item \cite{zerosum-globerson-2006} studied a two player game
    where the adversary deleted features that led to the maximum drop
    of a classifier accuracy and the defender built a robust
    classifier against malicious feature removal. The minimax
    solution was proposed with SVM as the underlying classifier. 
   In \cite{zerosum-Dekel-2010}, missing or corrupted
   features caused by an adversary can vary from one instance to another. 

  \end{itemize}  

\item A Stackelberg game where two players $S_L$ and $S_F$ take
  sequential actions is formulated as follows. Let $S_L$ be the
  leader with utility $U_L$. Let $S_F$ be
  the follower with utility $U_F$. The leader
  $S_L$ chooses a strategy $l$ first. After observing the leader's
  action, the follower $S_F$ chooses its strategy $f$. The game's subgame
  perfect equilibrium $(l^e,f^e)$ can be written as follows. 
$$
f_l = \textrm{argmax}_f ~S_F(l,f), ~~~(l^e,f^e) = \textrm{argmax}_l~ S_L(l,f_l).
$$
  Compared with zero-sum game and minimax
  solution, a Stackelberg game allows each player to maximize its
  own utility and offers a less conservative solution. 

  \begin{itemize}
 
  \item \cite{xi-game-2011} modeled adversarial classification as
    a Stackelberg game where the defender was the follower in the
    game. The adversary first transformed the objects under its 
    control knowing the action reduced the utilities of the
    malicious objects. Then the defender built a classifier
    minimizing the overall misclassification cost. For continuous variables,
    an approximate equilibrium solution was found by
    discretizing the variables and using a linear programming
    approach. A classifier's equilibrium performance was examined
    which indicated its long term success or failure facing
    a malicious adversary. 

 \item \cite{stackelberg-liu-2010} modeled the interaction
    between the defender and the adversary as a constant-sum Stackelberg game
    with convex loss. The equilibrium was the solution of a
    maxmin problem, solved via trust region methods. The game was
    repeated with the adversary 
    using different data transformations and the defender
    adjusting the classification rule.  

  \item \cite{stackelberg-bruckner-2011} let the defender be the
    leader in a Stackelberg prediction game. The equilibrium was
    the solution of a bilevel optimization problem, which can be
    solved by sequential quadratic
    programming. \cite{stackelberg-bruckner-2011} showed that the
    Stackelberg game with the worst case loss given the defender
    chose a hinge loss led to a SVM for invariances, and a Stackelberg
    game with linear loss led to a regular SVM. On the other hand,
    a Stackelberg game with logistic loss was the most robust
    against attacks.  

  \item \cite{stackelberg-zhou-2016} studied a  
single-leader-multiple-follower nested Stackelberg game with one defender
and  different types of adversaries. The game was formulated as 
multiple lower level Stackelberg games and one upper level
Bayesian Stackelberg game. The defender played a mixed
equilibrium strategy, which can be found by solving multiple
single-leader-single-follower games with probabilities determined
by the Bayesian Stackelberg game. Comparable to a mixed equilibrium
strategy, \cite{game-random-biggio-2017} studied randomized
defense strategy based on a game-theoretic formulation. 
  \end{itemize}  

\end{itemize}

\subsection{  Privacy Preserving Machine Learning}
\label{sec:privacy}

Privacy preserving mechanisms are developed to prevent privacy
leakage directly from a database or through a learning model. 
There are two main privacy preserving mechanisms, $k-$anonymity
\cite{k1,k2,k3} and differential privacy
\cite{dwork1,dwork2,dwork3}. $k-$anonymity directly perturbs the
individual data points, whereas differential privacy injects
noises to query results from a database. Two privacy preserving
mechanisms can be considered as complimentary to each other, as
suggested in \cite{clifton2013-k-diffpriv}. 
\cite{privacy-survey-2008} provided a survey of
  $k-$anonymity and distributed privacy preserving techniques
  and their application to data mining tasks.
\cite{privacy-publishing-survey2010} focused on various privacy
models for publishing privacy protected data records, not the
results from data mining or machine learning algorithms. Both
$k-$anonymity and differential privacy were discussed among other
approaches. Furthermore \cite{diffprivacy-stat-2010} focused on applying
differential privacy to statistics, such as statistical inference
and robust statistics. Below is a brief summary of $k-$anonymity
and differential privacy.

\begin{itemize}
\item A database satisfies
$k-$anonymity if every unique tuple for every combination of the variables
appears at least $k$ times. 

\item Differential privacy is a more
sophisticated mechanism. Let $Q$ be a randomized function used to
release information from a database. $Q$ satisfies $\epsilon-$
differential privacy if for two databases $D$ and $D'$ differing
by one record, 
$$
\frac{Pr(Q(D)\in H)}{Pr(Q(D')\in H)} \leq e^{\epsilon},~~~\forall~
H\in \textrm{range}(Q). 
$$    
The Laplace mechanism adds a Laplace noise to a query function $q$ 
to generate the randomized query result satisfying $\epsilon-$
differential privacy. An important concept to determine the size
of the Laplace noise is the sensitivity of the query function $q$. $q$'s
sensitivity equals to the maximum change of $q$ values over two
databases differing by one record, $D$ and $D'$. For Laplace
mechanism, the Laplace noise added query output is
$$
q(D)+\textrm{Laplace}\left({\frac{\textrm{sensitivity}}{\epsilon}}\right), 
$$
where $\epsilon$ is a pre-determined parameter (and hence the name $\epsilon-$
differential privacy). The recommended values for $\epsilon$ vary
in a big interval, from as small as 0.01 and 0.1 to as big as 7,
e.g., \cite{clifton-diffpriv-epsilon-2011}. 
\end{itemize}

Differential privacy
has been used to provide privacy 
guarantee to machine learning models. For example, 
\cite{dp-classifier-1,dp-classifier-2} produced privacy
preserving decision trees; \cite{dp-classifier-3} produced
privacy preserving SVM; \cite{dp-classifier-4,dp-classifier-5}
introduced privacy guarantee to deep learning. 
However, a recent membership inference attack against differentially
private deep learning model
\cite{member-infer-attack-diffprivateDNN-2018} 
suggested more research needs to be done to prevent privacy
leakage from complex learning models such as DNNs. 
There are several efforts along this line. 
\cite{member-infer-defense-2018} developed a min-max game and an
adversarial training procedure to protect DNNs from membership
inference attacks. Similarly
\cite{member-infer-defense-party-2018} also developed an
adversarial training procedure to defend against membership
inference attacks when multiple parties shared their data. 
\cite{protect-model-as-data-2018} mentioned the possibility of
extending data protection law to machine learning models by
considering them as personal data in legal terms, which pointed 
to an ultimate solution through regulation process. 

\subsection{  Defense Against Poisoning Attacks}
\label{sec:defense-poison}

To remove poisonous attack instances from the training data is a
difficult task. Here we discuss all the existing approaches proposed
to defend against poisoning attacks. We need more systematic
approaches to evaluate the impact of poisoning attacks and
mitigate the threat on learning techniques.   

\begin{itemize} 

\item \cite{poison-spam-2008} proposed Reject On Negative Impact (RONI)
defense and dynamic threshold defense. The idea behind RONI was
similar to that for outlier detection in linear regression. The
impact of every training email was measured as the difference in
performance by including and excluding it in the training
process. The training emails with large negative impact were
removed from the training data. However RONI did not perform well in
targeted poisoning attacks. Meanwhile, dynamic threshold defense recommended
dynamically adjusting threshold values in
SpamBayes. Although this approach increased accuracy with
legitimate emails, it had difficulty to correctly label spam emails. 

\item Facing poisoning attacks against principal component analysis
(PCA) subspace anomaly detection method in backbone network,
\cite{poison-pca-anomaly-2009} proposed a robust PCA based
defense approach. It maximized Median Absolute
Deviation (MAD) instead of variance to compute principal
components, and used a robust threshold value based on Laplace
distribution instead of Gaussian. \cite{poison-autoencoder-2018}
built an autoencoder based intrusion detection system, assuming
malicious poisoning attack instances were under 2\%. It demonstrated that the autoencoder
was more robust compared with a PCA based one.    

\item \cite{data-sanitizing-poison-2008} proposed a training data
sanitization scheme. The training data was broken into several
subsets, where subsets that belonged to different
networks or domains yielded good results in the sanitization
process. Each subset was used to train a model. These models were
used to label every training data point. 
A voting scheme was used to
determine whether a training data point was an attack instance or
not. Potential attack instances were then 
removed from training data. 

\item \cite{defense-poison-bagging-2011} argued poisoning attack
instances can be viewed as a special type of outliers. Because 
bagging can reduce the impact of outliers in training,
bagging ensembles constructed for spam filtering and intrusion
detection offered promising results. 

\item \cite{defense-poison-2017} analyzed the upper bounds on the loss
caused by poisoning attacks against SVM. Outliers were defined as
points far from cluster centroids, which were computed with or
without the poisonous points. An empirical online learning
algorithm was developed to compute the upper bound on the worst
case loss for any given dataset.  

\item \cite{poison-defense-activationcluster-2019} proposed activation
clustering method to detect the backdoor trigger in the training
data. The method examined  
activation of the last hidden layer in a DNN, where  
the poisonous samples containing the backdoor trigger appeared
in a separate cluster. 
\end{itemize}

\subsection{Robust DNN}
\label{sec:defense-evasion}

Unfortunately robustifying DNN proves to be a much more difficult
task compared to finding more effective adversarial
samples. Defense strategies were quickly found to be
vulnerable to newer and stronger attacks. Here we discuss the
main defense strategies proposed in the literature. 

\begin{itemize}
\item {\bf Adversarial Training}: The main idea behind
  adversarial training is to further regularize a DNN and
  improve its robustness. The procedure includes adversarial samples in 
  training, and continuously generating new
  adversarial samples at every step of training
  \cite{attack-dnn-1st-2014,advl-train-2015,attack-fgsm2-Kurakin-2017,advl-train-ensemble-2018}. \cite{attack-fgsm2-Kurakin-2017}
  reported ``label leaking'' for the adversarially trained
  network -- it is more accurate with the adversarial images
  generated using the same attack algorithm than
  the clean images. An improved procedure, ensemble adversarial
  training, used adversarial samples generated against other
  pre-trained models to decouple the attack algorithm and the
  model under attack. \cite{defense-gaussian-dataaugmentation-2017} proposed
  to enlarge the training data by adding Gaussian noise to clean
  images, which is computationally less expensive than using
  adversarial samples.    

\item {\bf Distillation}: \cite{distill-nn-2016} proposed
  defensive distillation using two
  networks. The first network was trained
  on the original training data. The second network was trained
  on data with the soft labels produced from
  the first network instead of the original labels.  
Meanwhile, \cite{distill-BayesNN-2018} showed experimentally Bayesian neural
networks (BNN) and distilled BNN accurately detected adversarial
samples if adversarial perturbations were generated with respect
to only one posterior sample of the network weights. 

\item {\bf Gradient Masking}: Gradient masking is another popular
  approach to defend against adversarial samples, e.g., 
  \cite{defense-gradientmask-2015,defense-gradientmask-2018,defense-gradientmask-nguyen2018,defense-gradientmask-ross2018,detect-2017-failed2}. There
  are three main gradient masking approaches: 1) A
  defense can cause gradient shattering, where gradients do no
  exist or pointing to the wrong direction, to render gradient
  based attacks ineffective; 2) A randomized defense can cause the
  gradients to be randomized as well, and reduce the
  effectiveness of attacks; 3) A defense can cause vanishing or
  exploding gradients when it involves several iterations of
  network computations.  

\item {\bf Pre-processing}: Carefully designed pre-processing
  procedures were also proposed to mitigate the effect of 
  adversarial perturbations. \cite{defense-feature-squeeze-2018} proposed feature
squeezing via spatial smoothing or reducing pixel color bit
depth.  \cite{defense-transform-2018} discovered image
transformations, such as total variance minimization and image
quilting, helped to eliminate adversarial perturbations. 
\cite{defense-GAN-2018} proposed to use a GAN to denoise
adversarial samples before feeding them into a classifier. 
Meanwhile, several winning teams in the 2017 NIPS competition
\cite{nips2017competition} applied image denoising techniques to
remove adversarial perturbations.   

\item {\bf Detection}: There are many detection approaches
  proposed. For example, \cite{detect-2017-failed1} proposed using
  a subnetwork as a detector. \cite{detect-advl-2018} computed a
  confidence score to detect both adversarial samples and out of class samples. 
\cite{defense-pixel-2018} used statistical hypothesis tests to
detect adversarial samples that were different from the clean 
image distributions. 

\end{itemize}

At the same time, many of these approaches were showed to be ineffective. 
\cite{attack-carliniL2-2017} developed the C\&W
attack specifically to break distillation defense. 
\cite{attack-10detection-2017} showed 10 detection methods failed
the C\&W attack soon after they were
published. \cite{attack-hiddengradients-carlini2018} showed
gradient masking was still vulnerable to attacks. 
\cite{advl-train-2015} had an experiment to demonstrate that an
ensemble was not more robust against adversarial perturbations. 
\cite{advl-train-ensemble-2018} mentioned adversarial training was
vulnerable to black box attacks.
Since a robust learning model must withstand different types of
attacks, to build a robust DNN remains a major challenge as of today. 

\section{Conclusions}
\label{sec:conclude}

Generative Adversarial Networks (GAN) 
 (e.g., \cite{gan-2014,gan-energy-2017,wgan-2017}) 
is sometimes mentioned together with adversarial machine
learning. GAN is a system deploying two competing neural
networks, i.e., a generator and a discriminator, to produce ultra-high
dimensional samples such as images. GAN is used also for image super-resolution
(e.g., \cite{gan-superresolution-2017}), for generating 3D
objects \cite{3D-gan-2016}, for generating new molecules
\cite{gan-molecule-2019} etc. GAN is essentially a generative
model. It is an important research topic but follows a different
direction compared with adversarial machine learning, which
focuses on learning model vulnerabilities and
robustness. For the interested readers, a GitHub site 
\footnote{https://github.com/nightrome/really-awesome-gan (last
  verified on 03/01/2020)}
has a list of GAN papers.

Most of the existing research in adversarial machine learning
focuses on supervised learning. On the other hand, providing
labels for a large amount of data points or samples from the
newest attacks may require expensive human expertise and becomes a
significant bottleneck. 
How to identify adversarial
samples in unsupervised and weakly supervised scenarios needs to
receive more attention. Clustering techniques and active learning techniques
also need to be robustified facing adversaries, e.g.,
\cite{malware-cluster-2009,intrusion-cluster-2015,
  advl-active-document-2016,advl-active-2017-minor,xi-active-2019}.  
Meanwhile, it is important to quantify the
  robustness and accuracy trade-off for machine learning
  algorithms facing adversarial attacks. Although recent works started to
  propose certain robustness or uncertainty measures, a more
  in-depth study on the trade-off is needed to build resilient
  learning algorithms.

Compared with a large number of digital attacks, there are only a
handful of physical attacks, e.g., 
\cite{attack-face-2016,stop-sign-2018,3d-turtle-2018,attack-BIM-physical-2017,dolphin-attack-2017,bioattack-2020}. 
They pose a more dangerous threat for learning
techniques. Physical attacks either design physical objects that
cannot be recognized by learning models
\cite{stop-sign-2018,3d-turtle-2018,attack-BIM-physical-2017}, or
they create malicious data targeting a particular vulnerability
in a hardware device \cite{dolphin-attack-2017}.
\cite{bioattack-2020} designed a moving physical
  object to make an object detection system ``blind'' because of
  the motion of the object.
A successful defense against
physical attacks may require multiple sensor systems to work
together with robust learning techniques and more secure hardware
designs.  

We cannot stop adversaries from launching
unexpected new attacks. However our efforts on secure
and robust machine learning techniques can mitigate the damages
caused by adversarial objects and slow down the arms race between
adversaries and defenders. 

\section*{Funding Information}
This work is supported in part by ARO grant W911NF-17-1-0356,
Purdue CERIAS seed grant, and a grant by Northrop Grumman Corp.

\bibliography{xi-wirescs}

\end{document}